\documentclass{article}
\usepackage{amsmath}
\usepackage{hyperref}
\usepackage{graphicx}

\input{tcilatex}
\begin{document}

\title{A Case for Lorentzian Relativity}
\author{Daniel Shanahan* \\
18 Sommersea Drive, Raby Bay, Queensland, Australia, 4163}
\maketitle

\begin{abstract}
The Lorentz transformation (the LT) is explained by changes occurring in the
wave characteristics of matter as it changes inertial frame. This
explanation is akin to that favoured by Lorentz, but informed by later
insights, due primarily to de Broglie, regarding the underlying unity of
matter and radiation. To show the nature of these changes, a massive
particle is modelled as a standing wave in three dimensions. \ As the
particle moves, the standing wave becomes a travelling wave having two
factors. \ One is a carrier wave displaying the dilated frequency and
contracted ellipsoidal form described by the LT, while the other (identified
as the de Broglie wave) is a modulation defining the dephasing of the
carrier wave (and thus the failure of simultaneity) in the direction of
travel. \ The superluminality of the de Broglie wave is thus explained, as
are several other mysterious features of the optical behaviour of matter,
including the physical meaning of the Schr\"{o}dinger Eqn. and the relevance
to scattering processes of the de Broglie wave vector. \ Consideration is
given to what this Lorentzian approach to relativity might mean for the
possible existence of a preferred frame and the origin of the observed
Minkowski metric.
\end{abstract}

\section{Introduction}

For the student of physics, there comes a moment of intellectual pleasure as
he or she realizes for the first time how changes of length, time and
simultaneity conspire to preserve the observed speed of light. \ Yet
Einstein's theory \cite{einstein1} provides little understanding of how
Nature has contrived this apparent intermingling of space and time. \ 

The language of special relativity (SR) may leave the impression that the
Lorentz transformation (the LT) describes actual physical changes of space
and time. \ Thus we have Minkowski's confident prediction that,

\begin{quotation}
Henceforth, space by itself, and time by itself, are doomed to fade away
into mere shadows and only a kind of union of the two will preserve an
independent reality \cite{minkowski}.
\end{quotation}

The impression that the LT involves some physical transmutation of
"spacetime" might seem consistent with the change of that nature
contemplated in general relativity (GR). \ But in GR a change in the metric
affects in like manner all that occupies the region of space in question. \
In SR it is necessary to distinguish what actually changes from what is
merely "observed"\footnote{%
"Observed" is used here, in the sense conventional in SR, to denote not what
is "seen" at a particular instant, but what the observer considers to have
occurred at that instant. \ What is seen includes light from events that
occurred progressively earlier the further they were away, an effect here
irrelevant.} to change.

Consider two explorers, who we will call Buzz and Mary. \ They had been
travelling, in separate space ships, side by side, in the same direction. \
But Buzz has veered away to explore a distant asteroid. \ Mary concludes
from her knowledge of the LT that time must now be running more slowly for
Buzz and that he and his ship have become foreshortened in the direction
that Buzz is travelling relative to her. \ Buzz observes no such changes
either in himself or in his ship. \ To Buzz, it is in Mary and her ship that
these changes have occurred. \ Buzz is also aware that events that he might
previously have regarded as simultaneous are no longer so. \ 

No relevant physical change has occurred in Mary or her spacecraft. \ She
has not accelerated. \ She is in the same inertial frame as before. \ Nor
(ignoring gravitational effects, in this case negligible) has any actual, as
distinct from observed, change occurred in the space through which the
travellers are moving. \ To suggest otherwise would be to suppose that space
is able to contract in one way for one particle and in a different way for
another moving relatively to the first, albeit that the two (or at least
their correspondingly contracted fields) could be occupying the very same
piece of space. \ Even Buzz will have realized, as he observed the
contraction of the constellations, that the stars were not in fact closing
ranks around him.

A change of inertial frame \textit{has} occurred for Buzz and his
spacecraft. \ It must be this change that is the source of the changes\ that
the two explorers are observing. \ It is not difficult to understand that
Buzz's change of velocity may have caused a change in him that has affected
how he is perceived by Mary. \ But it must also be the case, since nothing
else has changed, that it is this same change in Buzz that has caused him to
consider the (in fact unchanged) Mary in a different light. \ 

Buzz will not sense that he has changed. \ After all everything in his
inertial frame will have changed in like manner. \ Unlike the carousel rider
who sees the fairground whirling about her, but is under no illusion as to
what is really happening, Buzz has suffered relativistic changes in his
vital processes, and lost the means of discernment. \ For Buzz the LT will
describe very well his altered perspective. \ But it would be as
inappropriate to \textit{explain} length contraction, time dilation and loss
of simultaneity as resulting from a physical transformation of space or
spacetime as it would be to describe the rotation of an object in 3-space as
a rotation \textit{of} space rather than a rotation \textit{in} space. \ 

While one might wish to elevate the discussion by reference to differential
manifolds, the spacetime continuum, or the Minkowski metric, the curious
effects described by the LT must be explained by a change that occurs in
matter as it suffers a "boost" from one inertial frame to another. \ Indeed
the Minkowski metric should itself be seen as a kind of illusion, and as a
consequence rather than the cause of this change in matter.

But to entertain these thoughts is to embark upon a process of reasoning,
associated primarily with Lorentz, that became unfashionable\footnote{%
Described as "best forgotten except by historians" by Rindler\textit{\ }\cite%
{rindler}, p 11. \ For views more consonant with those offered here, see
Bell\ \cite{bell}, Brown and Pooley \cite{brownpooley2}, and in particular,
Brown\textit{\ }\cite{brown}\textit{. \ }The pedagogical merits of this
"constructive" approach to SR are also discussed in Miller \cite{miller},
and Nelson \cite{nelson}.} following Einstein's famous paper of 1905 \cite%
{einstein1}. \ Lorentz had sought to explain the transformation that bears
his name on the basis of changes that occur in matter as it changes
velocity. \ This was, it is suggested, an idea before its time. \ We will
consider in this paper how Lorentz's program might have proceeded if
informed by later insights as to the underlying wave nature of matter,
including ironically those of Einstein himself, but in particular that of de
Broglie \cite{thesis}.

\section{Lorentz and Einstein}

Briefly first some history. \ The problem addressed by Lorentz and
subsequently Einstein was the speed of light. \ This emerged as a constant
in Maxwell's equations, but\ if as was generally supposed, light is
wave-like, it seemed reasonable to assume that it must be carried by some
medium (the "luminiferous aether") at a velocity characteristic of that
medium. \ Thus its velocity relative to an observer should have varied with
the motion of the observer through the medium. \ Experiments of increasing
sophistication failed to reveal any trace of that variation.

Several explanations were put forward. \ It was proposed that the Earth must
carry the local aether with it, but a more fruitful suggestion made
independently by Fitzgerald \cite{fitzgerald} and Lorentz \cite{lorentzfitz}
was that objects moving through the aether must be somehow shortened along
their direction of travel, thereby disguising relative changes in the
velocity of light. \ It was supposed that intermolecular forces must be
transmitted at the same velocity as electromagnetic waves, so that movement
through the aether would influence the degree of attraction between
molecules and thus the separation of those molecules. \ 

To effect a reconciliation with Maxwell's equations, it was necessary to
assume changes not only of length, but also of time, and thus the LT, 
\begin{align*}
x^{\prime }& =\gamma \left( x-vt\right) , \\
y^{^{\prime }}& =y, \\
z^{^{\prime }}& =z, \\
t^{\prime }& =\gamma \left( t-\frac{vx}{c^{2}}\right) ,
\end{align*}%
where $c$ is the speed of light, $\gamma $ is the Lorentz factor,%
\begin{equation}
\left( 1-\frac{v^{2}}{c^{2}}\right) ^{-\frac{1}{2}},  \label{lorfac}
\end{equation}%
and the unprimed coordinates are for an event in the observer's rest frame,
while the primed are for those in a frame moving in the $x$ direction at the
velocity $v$.

The LT was already reasonably well known by 1905. \ There had been
significant contributions to its development, not only from Lorentz and
Fitzgerald, but also by (among others) Heaviside, Larmor and Poincar\'{e}. \
It was Heaviside's analysis of the disposition of fields accompanying a
charged particle (the "Heaviside ellipsoid") that had suggested to
FitzGerald the idea of length contraction \cite{heaviside}. \ Larmor had
described an early form of the LT and discussed the necessity of time
dilation \cite{larmor}. \ Poincar\'{e} had recognized the relativity of
simultaneity and had studied the group theoretic properties that form the
basis for the covariance of the transformation \cite{poincare}.

But these "trailblazers" (Brown \cite{brown}, Ch. 4 ) appear to have missed
in varying degrees the full significance of the transformation\footnote{%
There are differing opinions as to who knew what in 1905. \ For an
interesting sampling of conflicting views on priority (Einstein, Lorentz or
Poincar\'{e}?) see the preamble to Reignier \cite{reignier}.}. \ It is not
only particular phenomena, but all of Nature that changes for the
accelerated observer. \ Lorentz struggled to explain how all aspects of
matter could became transformed in equal measure, being discouraged by
experimental reports that seemed to show that particles do not contract in
the direction of travel (see Brown \cite{brown}, p. 86). \ A wider view
seems to have been noticed by Poincar\'{e} \cite{poincare}, who has been
regarded by some as codiscoverer of SR (see, for instance, Zahar \cite{zahar}%
, and Reignier \cite{reignier}). \ But it is not apparent that these earlier
investigators saw the changes described by the LT as anything more than
mathematical constructs.\ \ In his paper of 1905 \cite{einstein1}, Einstein
simply asserted that the velocity of light, and other properties of Nature,
must appear the same for all uniformly moving observers, thereby effecting
an immediate reconciliation between Maxwell's equations and classical
mechanics.

In 1905, Einstein's approach may have been the only way forward. \ It was
not until 1924, only a few years before the death of Lorentz, and well after
that of Poincar\'{e}, that de Broglie proposed that matter is also wavelike 
\cite{thesis}, an idea that might have suggested to Lorentz why molecules
become transformed in the same degree as intermolecular forces. \ But as
inadequate and ad hoc as Lorentz's suggestions may have seemed at the time,
he at least had sought an underlying physical basis for the transformation.
\ Lorentz commented, presumably with some chagrin, that,

\begin{quotation}
Einstein simply postulates what we have deduced, with some difficulty, and
not altogether satisfactorily, from the fundamental equations of the
electromagnetic field \cite{lorentz2}. \ 
\end{quotation}

In what follows, the distinction drawn will be between Einstein's SR (ESR)
and what we will call Lorentzian SR (LSR). \ This is not to diminish the
contributions of others, but it was Lorentz in particular who sought to
explain SR from underlying physical processes, as will be the objective
below. \ Once the form of the LT was known, all else in SR then followed,
including the composition of velocities, the group theoretic properties of
the transformation, and the invariance of Maxwell's equations. \ It may be
argued that with these refinements (largely due to Einstein and Poincar\'{e}%
), ESR and LSR are essentially equivalent. \ They cannot be distinguished,
mathematically or empirically, through the privileged frame that was
supposed by Lorentz, but declared "superfluous" by Einstein \cite{einstein1}%
. \ It would seem that any such frame is rendered undetectable by the
covariance of the LT. \ Nor can ESR and LSR be distinguished by supposing
that in ESR, though not in LSR, the LT describes a transformation of
spacetime. \ As we have seen, the LT must be explained in either case by
changes occurring in matter as it is accelerated from one inertial frame to
another.

Historically, the two approaches are distinguished by the assumptions made
by Einstein in justifying the LT. \ His confident assertion of these
"postulates" gave impetus to the recognition and development of SR. \ But by
raising "to the status of a postulate", the conjecture that Nature displays
"no properties corresponding to the idea of absolute rest", and introducing
the further postulate that the speed of light must be the same for all
observers \cite{einstein1}, Einstein avoided all necessity of explaining how
Nature has arranged these matters. \ He left the impression that there was
nothing more that needed explaining.

Einstein's postulates presuppose that everything in Nature transforms in the
same degree. \ Accepting that this is so, we should suspect the existence of
some unifying feature common to matter and radiation that ensures that it is
so. \ If it had been possible to see the LT as a transformation of space
carrying with it all within space, this would have explained very nicely the
all-encompassing generality of the transformation. \ Having dismissed that
possibility, we will proceed now on a different tack.

\section{The scheme of the paper}

It will be supposed (taking our cue from de Broglie \cite{thesis}) that
radiation and matter transform in like manner because they are constituted
in like manner from similar wave-like influences. \ By modelling a massive
particle as a standing wave in three dimensions, it will be shown in the
next section (Sect. 4) that the changes in length, time and simultaneity
described by the LT are the immediate consequences of changes in the wave
structure of the particle as it changes inertial frame. \ 

As discussed in Sect. 5, a compelling advantage of this treatment will be
the emergence of the otherwise mysterious de Broglie wave, not as the
independent wave generally supposed, but as a modulation defining the
dephasing of the underlying wave structure in the direction of travel. \ \
This conception of the de Broglie wave is not itself new. \ Such a
modulation may be discerned in a toy model described by de Broglie himself
in his famous doctoral thesis of 1924 \cite{thesis}. \ It has been discussed
in various contexts on several occasions since [18-22]. \ However, it will
be shown here that this interpretation of the wave explains several curious
features of the wave-like behaviour of matter. \ It avoids the anomalous
nature of the superluminality\ of the wave, it makes sense of the role of
the wave in quantization and in the Schr\"{o}dinger Equation; and it
explains the relevance of the de Broglie wave number to the\ optical
properties of massive particles. \ Its primary significance to this paper is
that the dephasing defined by the modulation is the measure of the failure
of simultaneity that is perhaps the most counter-intuitive aspect of SR.

In Sect. 6, we will consider the sufficiency of the model particle
introduced in Sect. 4. \ It will be shown that this test particle exhibits,
not so much the properties of all matter, but more relevantly, those
implicitly assumed by ESR. \ In Sect. 7, consideration will be given to the
possible existence of a preferred frame, that is to say, a frame in which
the speed of light is not simply observed to be, but in fact is, the same in
all directions. \ The position will be taken that the fixed rate at which
photons are observed to propagate through space suggests very strongly, if
not a light-carrying medium, then at least the preferred frame favoured by
Lorentz.

In Sect. 8, it will be argued that "Lorentz invariance" - the invariance of
the laws of physics to all inertial observers - is itself a consequence of
the wave-like characteristics of matter. \ A brief summary in\ Sect 9 will
conclude the paper.

\section{A model particle}

In 1905,\ Einstein took as his yardstick, the rigid rod, avoiding
consideration of underlying structure \cite{einstein1}. \ However a massive
particle has, from the Planck-Einstein relation,%
\begin{equation}
E=\hbar \omega _{E}=\hbar \gamma \omega _{o},  \label{planck}
\end{equation}%
an associated frequency $\omega _{E}$, and from the de Broglie relation,

\begin{equation}
p=\hbar \kappa _{dB}=\hbar \gamma \kappa _{o}\frac{v}{c},  \label{deb}
\end{equation}%
a wave number $\kappa _{dB}$ (the de Broglie wave number), where $E$ and $p$
are respectively the energy and momentum of the particle, $\hbar $ is the
reduced Planck's constant, and $\omega _{o}$ is the natural or
characteristic frequency of the particle at rest. \ 

Wave number $\kappa _{dB}$ and frequency $\omega _{E}$ define the wave that
de Broglie referred to as a pilot wave, and that we now know as the de
Broglie or matter wave. \ Considered as an independent wave it is anomalous.
\ It has in free space the form of a transverse plane wave,%
\begin{equation*}
e^{i(\omega _{E}t-\kappa _{dB}x)},
\end{equation*}%
of superluminal velocity $c^{2}/v$. \ What is also curious (though a pointer
to the true nature of this wave) is that its speed increases as the particle
slows, becoming infinite as the particle comes to rest.

\includegraphics[width=8.0cm]{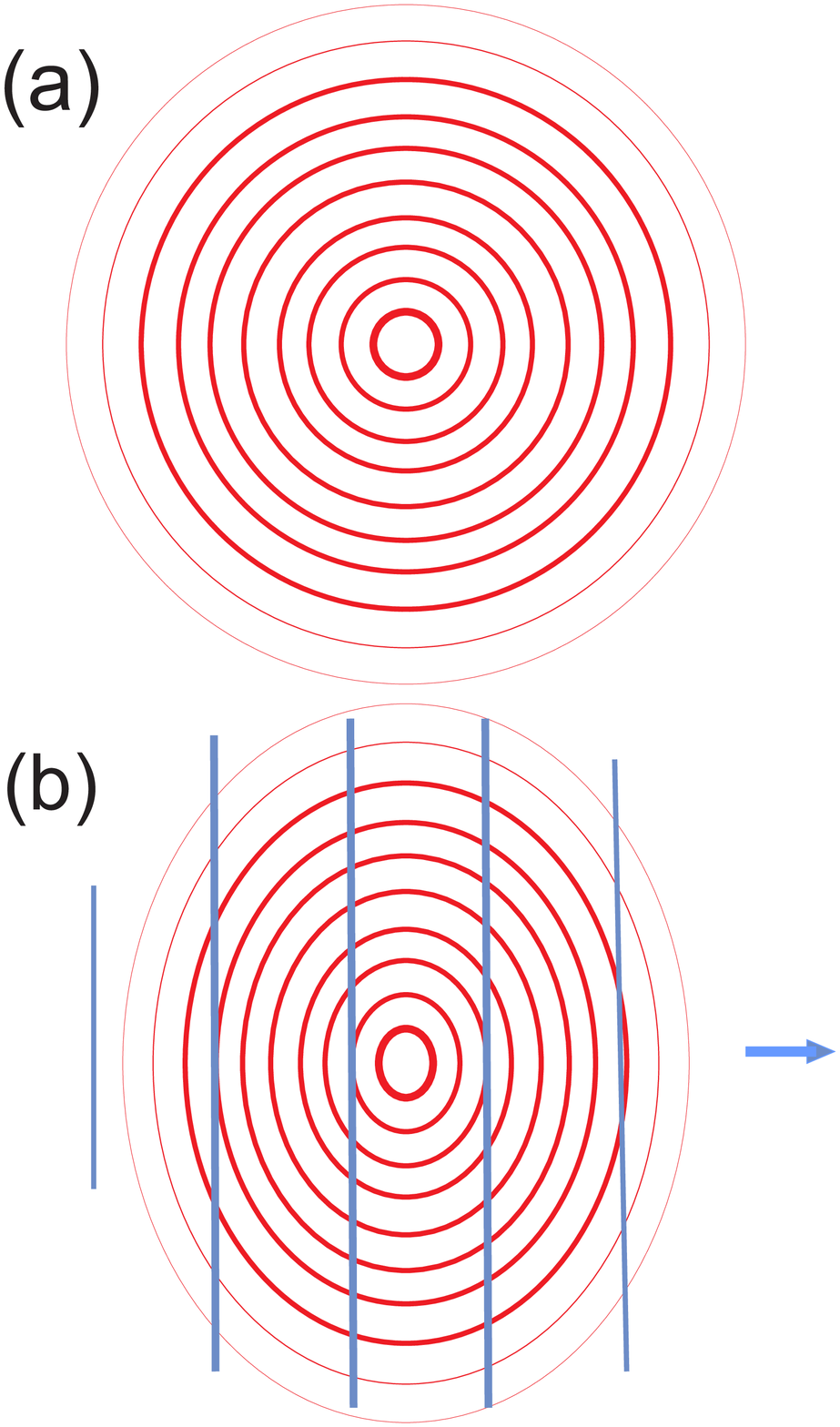}

\begin{quotation}
\textbf{Fig. 1} \ A model particle is represented: (a) as a spherical
standing wave; and (b) as a relativistically contracted carrier wave of
velocity $v$, subject to a modulation (the de Broglie wave) of superluminal
velocity $c^{2}/v$. \ The ellipses in (b) represent the wave fronts of the
carrier wave and the vertical lines those of the de Broglie
modulation.\medskip \medskip
\end{quotation}

In its rest frame, the particle retains nonetheless the frequency $\omega
_{o}$, which we will assume is not a fictitious or merely internal
frequency, but the measure of some oscillatory disturbance communicated
through space at velocity $c$. \ From this assumption we model a massive
particle in its rest frame by the simple symmetrical waveform, 
\begin{equation*}
\Psi \left( \mathbf{r},t\right) =\left\vert \mathbf{r}\right\vert
^{-1}[e^{i\left( \omega _{o}t-\mathbf{\kappa }_{o}\cdot \mathbf{r}\right)
}-e^{i\left( \omega _{o}t+\mathbf{\kappa }_{o}\cdot \mathbf{r}\right) }]/2,
\end{equation*}%
or, taking real parts, 
\begin{equation}
\Psi \left( \mathbf{r},t\right) =\left\vert \mathbf{r}\right\vert ^{-1}\sin 
\mathbf{\kappa }_{o}\cdot \mathbf{r\,\,}\cos \omega _{o}t,  \label{model}
\end{equation}%
which is a spherical standing wave centred at $\mathbf{r}=0$ and comprising
incoming and outgoing waves of velocity $c$, where,%
\begin{equation*}
\frac{\omega _{o}}{\kappa _{o}}=c,
\end{equation*}%
so that $\kappa _{o}$ is not here the de Broglie wave number but the wave
number that must be associated with a wave of frequency $\omega _{o}$ and
velocity $c$\footnote{$\kappa _{o}$ also has significance as the reduced
Compton wave number.}.

Model particle (\ref{model}) is depicted (in two dimensions) in Fig. 1(a). \
This particular structure is evidently unphysical because of the singularity
at $r=0$. \ It is nonetheless the simplest expression of the assumption,
essential to ESR, that influences moving to and from (and indeed through) a
particle do so at the speed of light. \ We will consider this model further
in Sect. 6. \ We now investigate how this particle must change if it is to
acquire a velocity $v$ in the frame\ of the laboratory, when subject to the
constraint that its constituent influences (rays) must retain the velocity $%
c $ with respect to that frame. \ (We are thus assuming for the moment that
the inertial frame in which this particle has the form (\ref{model}), which
we have called the laboratory frame, is at the same time the preferred or
privileged frame supposed by Lorentz.\ \ Of course, we do not know and, as
discussed in Sects. 7 and 8, likely cannot\ not know, the actual location of
this frame. \ To simplify matters further the amplitude $\left\vert \mathbf{r%
}\right\vert ^{-1}$ is omitted in what follows).

What we require now is not a standing wave but a travelling wave. \ Its form
can be established in either of two ways, and it will be instructive to
consider both. \ The first is by construction, and we begin by considering
rays, directed forward and rearward along the direction of travel, which we
will take to be the positive $x$-direction. \ At rest in the preferred
(laboratory) frame, the composition of these rays results in the
one-dimensional standing wave

\begin{align}
\Psi (x,t)& =[e^{i(\omega _{o}t-\kappa _{o}x)}-e^{i(\omega _{o}t+\kappa
_{o}x)}]/2,  \notag \\
& =\sin \kappa _{o}x\cos \omega _{o}t.  \label{swave}
\end{align}

\bigskip

This standing wave becomes a travelling wave of velocity $v$ if the wave
characteristics of the rays directed forward and rearward (which we now
label 1 and 2 respectively) become,

\begin{align}
\omega _{1}& =\gamma \omega _{0}(1+\frac{v}{c}),\;\;\omega _{2}=\gamma
\omega _{0}(1-\frac{v}{c}),  \notag \\
\kappa _{1}& =\gamma \kappa _{0}(1+\frac{v}{c}),\;\;\kappa _{2}=\gamma
\kappa _{0}(1-\frac{v}{c}),  \label{conv}
\end{align}%
where $\gamma $ is again the Lorentz factor (\ref{lorfac}), and since,%
\begin{equation}
\frac{\omega _{1}}{\kappa _{1}}=\frac{\omega _{2}}{\kappa _{2}}=\frac{\omega
_{0}}{\kappa _{0}}=c,  \label{wk}
\end{equation}%
both rays retain as required the velocity $c$ with respect to the preferred
frame. \ We then have by composition, 
\begin{align}
\Psi \left( x,t\right) & =[e^{i\left( \omega _{1}t-\kappa _{1}x\right)
}-e^{i\left( \omega _{2}t+\kappa _{2}x\right) }]/2,  \label{twave1} \\
& =\sin (\frac{\omega _{1}-\omega _{2}}{2}t-\frac{\kappa _{1}+\kappa _{2}}{2}%
x)\cos (\frac{\omega _{1}+\omega _{2}}{2}t-\frac{\kappa _{1}-\kappa _{2}}{2}%
x),  \label{twave2}
\end{align}%
\newline
which from relations (\ref{conv}) may also be written,%
\begin{equation}
\Psi \left( x,t\right) =\sin \gamma (\omega _{0}vt/c-\kappa _{0}x)\cos
\gamma (\omega _{0}t-\kappa _{0}vx/c),  \label{oned2}
\end{equation}%
and is a travelling wave of the kind illustrated in Fig. 1(b).\ \ This
one-dimensional travelling wave already displays features that will become
more apparent when we consider the full three-dimensional travelling wave,
namely a carrier wave of velocity $v$ (the first factor in Eqn. (\ref{oned2}%
) ) and a superluminal modulation (a beating) of velocity $c^{2}/v$ (the
second factor)\footnote{%
The effect is strictly a "beating" between interfering waves of equal
amplitude but differing characteristics. \ We describe the wave factors as
"carrier" and "modulation", but which is carrier and which is modulation is
somewhat arbitrary.}

\includegraphics[width=8.0cm]{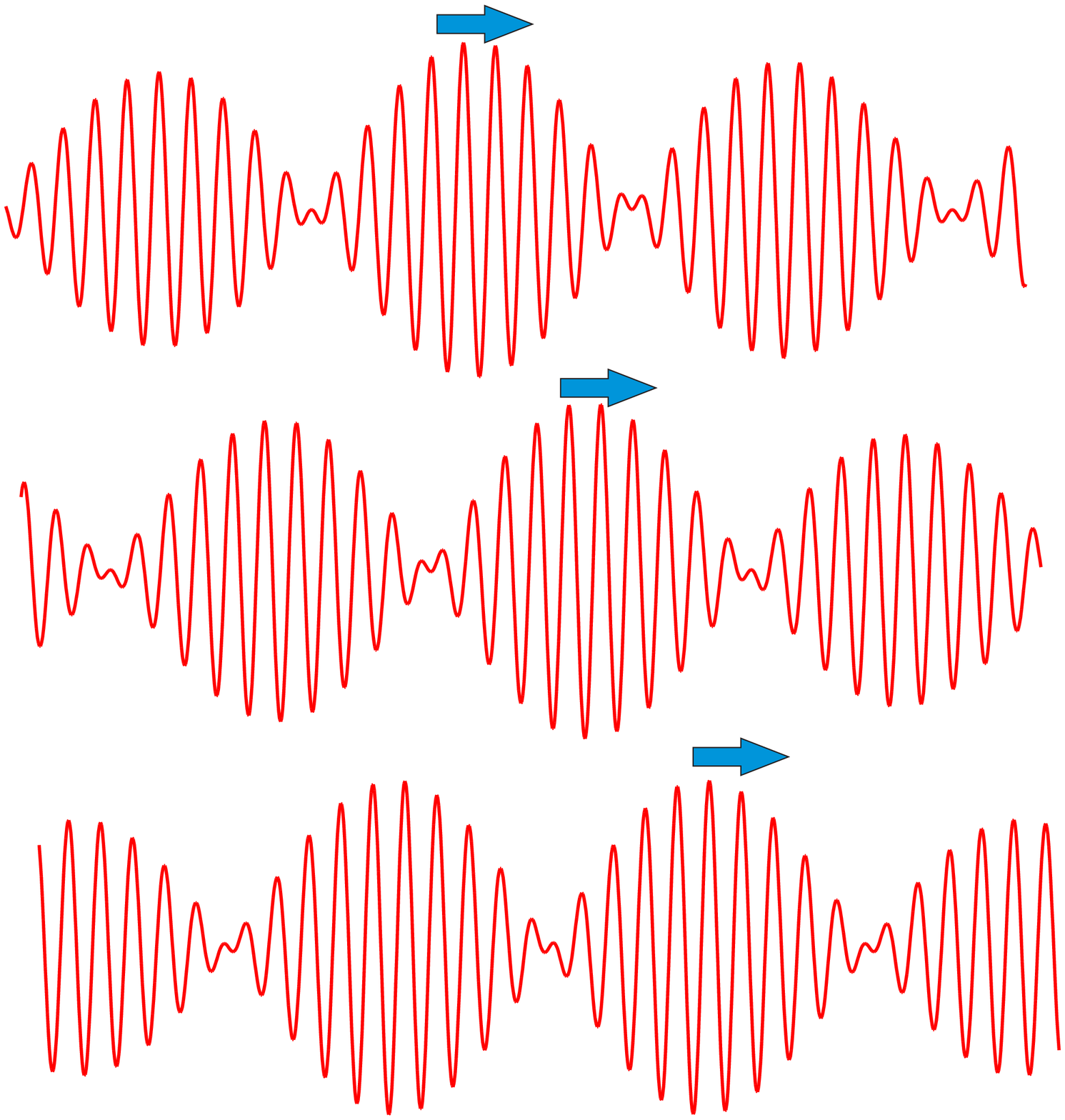}

\begin{quotation}
\textbf{Fig. 2} \ \ Amplitudes (indicative only) at times $t_{1}<t_{2}<t_{3}$
for a wave of the kind shown in Fig. 1(b). \ As shown by the arrows, the
modulation (the de Broglie wave) develops through the carrier wave. \ It
moves at the superluminal velocity $c^{2}/v$, while the carrier wave moves
at velocity $v$, the classical velocity of the particle.\medskip \medskip 
\end{quotation}

The structure of the full model wave at velocity $v$ is now obtained by
noticing that (as shown in Fig. 3) the amplitude of this wave at any point $%
P $ at time $t$, when the centre of the wave has reached $B$, results from
the interference of the outgoing ray that left the particle centre when it
was at $A$ at the earlier time $t-t_{1}$, with the incoming ray that will
reach the particle centre when it is at $C$ at the later time $t+t_{2}$. \
(To an observer in the frame of the moving particle, the paths of these rays
will appear to coincide, but they do not coincide in the laboratory frame).

\includegraphics[width=8.0cm]{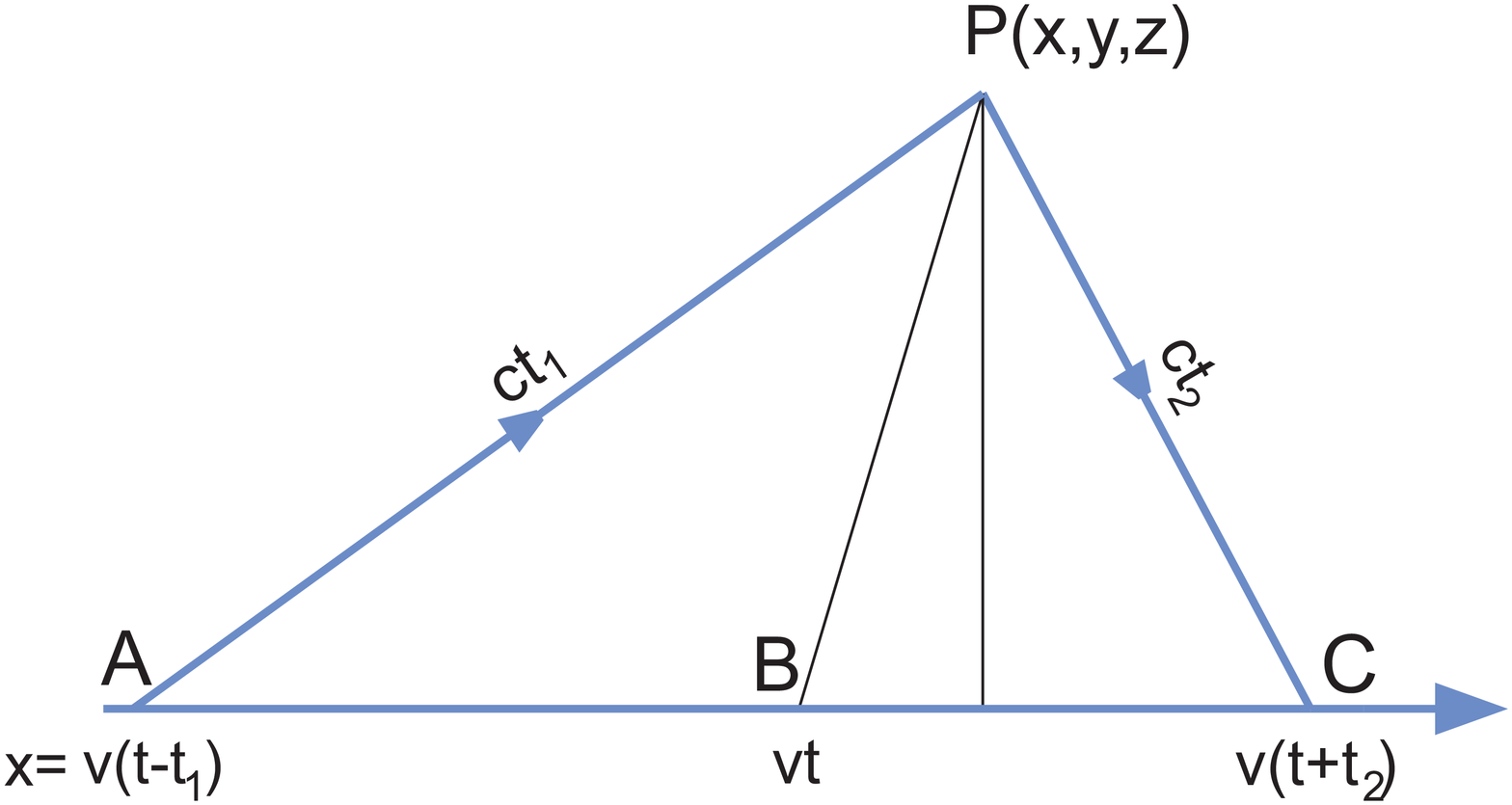}

\begin{quotation}
\textbf{Fig. 3} \ \ With the centre of the wave of Fig. 1(b) at $B$,
interference occurs at $P(x,y,z)$ between the outgoing ray that left the
wave centre at $A$ at the earlier time $t-t_{1}$, and the incoming ray that
will not reach the centre until it is at $C$ at the later time $t+t_{2}$. \
The waveform thus constructed corresponds exactly to that obtained by
Lorentz transforming the wave of Fig. 1(a).\medskip \medskip
\end{quotation}

In a more realistic model, phase might be expected to vary with angular
displacement about the particle centre, but in this simple model we have
supposed that all rays passing through the centre are of the same phase. \
Thus, from Eqn. (\ref{twave2}) for the one dimensional case,%
\begin{align}
\Psi _{A}(t-t_{1})& =\exp i\left[ \frac{\omega _{1}+\omega _{2}}{2}(1-\frac{%
v^{2}}{c^{2}})(t-t_{1})\right] ,\;and  \label{phaseA} \\
\Psi _{C}(t+t_{2})& =\exp i\left[ \frac{\omega _{1}+\omega _{2}}{2}(1-\frac{%
v^{2}}{c^{2}})(t+t_{2})\right] ,  \label{phaseB}
\end{align}%
(where the arguments of $\Psi _{A}$ and $\Psi _{C}$ are the phases at $A$
and $C$ respectively).

We now need $t_{1}$ and $t_{2}$. \ From Fig. 3,%
\begin{equation*}
c^{2}t_{1}^{2}=[(x-v(t-t_{1})]^{2}+y^{2}+z^{2},
\end{equation*}

from which,%
\begin{equation}
t_{1}=\frac{\frac{v}{c^{2}}(x-vt)+\frac{1}{c}\sqrt{(x-vt)^{2}+\left(
1-v^{2}/c^{2}\right) \left( y^{2}+z^{2}\right) }}{1-v^{2}/c^{2}},  \label{t1}
\end{equation}

and similarly,%
\begin{equation}
t_{2}=\frac{-\frac{v}{c^{2}}(x-vt)+\frac{1}{c}\sqrt{(x-vt)^{2}+\left(
1-v^{2}/c^{2}\right) \left( y^{2}+z^{2}\right) }}{1-v^{2}/c^{2}}.  \label{t2}
\end{equation}

Composing Eqns. (\ref{phaseA}) and (\ref{phaseB}), and substituting for $%
t_{1}$ and $t_{2}$ from Eqns. (\ref{t1}) and (\ref{t2}) respectively, we
obtain after some algebra the full travelling wave,

\begin{equation}
\Psi \left( x,y,z,t\right) =\sin \kappa _{o}\sqrt{\gamma
^{2}(x-vt)^{2}+y^{2}+z^{2}}\cos \gamma \omega _{o}(t-vx/c^{2}),
\label{moving}
\end{equation}%
which is sketched for a particle of relativistic velocity in Fig. 1(b). \
The corresponding amplitudes of such a wave (in the $x$-direction) are of
the general form suggested by Fig. 2.

Travelling wave (\ref{moving}) could have been obtained by simply Lorentz
transforming model wave (\ref{model}). \ In following the longer route we
have not derived the LT, for we have assumed the Lorentz factor $\gamma $
(an assumption that will be examined in Sect. 8), but we have demonstrated
the physical effect of the LT, not as Einstein did with reference to rigid
rods and synchronous clocks, but as Lorentz might have done, from the
underlying oscillatory nature of matter. \ In doing so we have of course had
the advantage of a unified view of matter and radiation that was not
available to Lorentz.

It is important to notice the composite form of the transformed wave (\ref%
{moving}). \ It has two factors. \ The first, 
\begin{equation}
\sin \kappa _{o}\sqrt{\gamma ^{2}(x-vt)^{2}+y^{2}+z^{2}},  \label{Lorentz}
\end{equation}%
is a form of carrier wave. \ It moves with velocity $v$, the classical
velocity of the particle and, as suggested by the ellipses in Fig. 1(b),
describes the relativistically contracted ellipsoidal form of the moving
particle as considered from the standpoint of an observer in the laboratory
frame. \ It has the frequency $\omega _{o}/\gamma $. \ The second factor,%
\begin{equation}
\cos \gamma \omega _{o}(t-vx/c^{2}),  \label{de Broglie}
\end{equation}%
is a plane wave of superluminal velocity $c^{2}/v$, the wave fronts of which
are suggested by the vertical lines in Fig. 1(b). \ This plane wave has the
wave number and frequency of the de Broglie wave, and is indeed the de
Broglie wave. \ But it is not here an independent wave but a modulation of
the carrier wave (\ref{Lorentz}), describing the dephasing of that wave (and
thus the failure of simultaneity) in the direction of travel of the particle.

Travelling wave (\ref{moving}) displays all the effects predicted by SR -
length contraction, time dilation and failure of simultaneity. \ If all
matter changed between inertial frames in the same manner as our model
particle, this would explain why Mary considered Buzz to have changed in
accordance with the LT. \ But the transformation is here a change of wave
structure, not of space, nor even yet of reference frame. \ There has been
no opportunity, let alone necessity, for some additional transformation of
"spacetime". \ 

What, then, of Buzz's belief that it was Mary who changed. \ In its altered
state, the model particle has become a somewhat untidy affair. \ But to an
observer moving with the particle (such as Buzz), and transformed in like
manner, the particle will seem to have its original form (\ref{model}). \
Being composed of influences developing at the speed of light, this observer
will also lack the means of discerning differences in his or her velocity
relative to that of light. \ In whatever inertial frame the observer
occupies, the covariance and group theoretic properties of the LT will
guarantee the constancy of the observed speed of light and the laws of
physics. \ It is these properties that ensure that if a preferred frame does
exist (of which more will be said in Sect. 7) it is likely to remain
undetected.

\section{The de Broglie wave}

Once the de Broglie wave is seen as a modulation rather than an independent
wave, several mysteries become resolved. \ As a modulation, its superluminal
velocity is no longer that of energy transport and need not be explained
away by the usual but awkward device of equating the velocity of the
particle with the group velocity of a packet of such de Broglie waves. \ It
is also only natural that the velocity of this modulation should increase as
the particle slows, and become infinite as the particle comes to rest. \ At
rest, the crests of the composite wave are no longer peaking in sequence,
but in unison. \ Simultaneity has been restored, alignment of phase has
become instantaneous, and the velocity of the modulation describing the
progress of that alignment has thus become infinite. \ In effect the de
Broglie wave now disappears.

Other difficulties are explained. \ It can be seen\ why the de Broglie wave
cannot be fitted to electron orbits. \ It is not the superluminal modulation
but the full composite wave that follows the orbital path. \ Yet it \textit{%
is} the modulation that defines the phase lost by the electron in following
that path, and it is thus the de Broglie wave number $\kappa _{dB}$ of Eqn. (%
\ref{deb}) that appears in Bohr quantization conditions for stable orbits,
which are of general form,

\begin{equation}
\oint\kappa_{dB}\,ds=2n\pi\;\;\;\;(n=1,2,3...),  \label{bohr}
\end{equation}
a requirement that evidently ensures continuity in the wave structure
generated by the orbit in question.

The\ Schr\"{o}dinger equation (the SE) is also based on the modulation
rather than the full wave, thus explaining the difficulties that have been
experienced in according physical significance to solutions of this equation%
\footnote{%
Of Schr\"{o}dinger's own difficulties in interpreting those solutions, see
Dorling \cite{dorling}.}. \ In constructing a wave equation that would have
solutions consistent with the Planck-Einstein and de Broglie relations
(Eqns. (\ref{planck}) and (\ref{deb})\ above), Schr\"{o}dinger made the
substitutions,%
\begin{align*}
p& \rightarrow i\hbar \frac{\partial }{\partial x},\;and \\
E& \rightarrow i\hbar \frac{\partial }{\partial t},
\end{align*}%
in the non-relativistic equation of motion, 
\begin{equation*}
E^{2}=\frac{p^{2}}{2}+V,
\end{equation*}%
to obtain the non-relativistic SE,%
\begin{equation*}
i\hbar \frac{\partial \psi }{\partial t}=-\frac{\hbar ^{2}}{2m}\nabla
^{2}\psi +V\psi ,
\end{equation*}%
and likewise in the relativistic equation of motion to obtain the
corresponding wave equation (now called the Klein-Gordon equation).

But as we have seen, Eqns. (\ref{planck}) and (\ref{deb}) define the wave
characteristics, not of the full waveform, but of the de Broglie wave. \
Thus solutions to the SE, whether in relativistic or non-relativistic form,
are not actual waves but distorted mappings of a modulation. \ These\
solutions are able to identify allowed energy levels only because each such
solution is homologous to a set of quantization conditions, the continuity
implied by those conditions being at the same time the continuity required
for physically realistic solutions of the SE.

Considered as a modulation, it also becomes understandable, not only that
the superluminal wave does not outrun the subluminal particle, but why the
wave vector $\mathbf{\kappa }_{dB}$ of this wave has a controlling influence
on the optical properties of the particle. \ Any change in the trajectory of
the particle will be accompanied by a rotation throughout space of the wave
fronts of the de Broglie wave and, as the immediate consequence of that
rotation, a rearrangement of phase (and simultaneity) throughout the entire
composite wave structure. \ It is only necessary to assume that a scattered
particle (for instance a diffracted particle) will prefer a trajectory that
preserves its characteristic transverse wave form to see why the particle
tends to adopt a path through the scattering medium in which the de Broglie
wave recombines coherently, that is to say, by which its recombining parts
interfere constructively rather than destructively. \ 

\section{The sufficiency of the model}

It might be objected that without knowing all there is to know about the
constituent particles and fundamental forces of Nature, we are no better
equipped than was Lorentz to base the LT on the properties of matter. \ It
might also be argued that what we have been assuming are waves are not waves
at all but particles that display oscillatory characteristics, or if waves,
not real waves, but the probabilistic waves of standard quantum mechanics
(SQM).

The model assumes that a massive particle comprises influences of frequency $%
\omega _{o}$ moving to and from the particle centre at velocity $c$. \ That
the interaction of a particle with its fellows can be expressed in terms of
incoming and outgoing influences is obvious and well recognized\footnote{%
See, for example, the discussions of \ advanced and retarded influences by
Wheeler and Feynman \cite{wheeler1} \cite{wheeler2}. \ Those papers
concerned action at a distance but such incoming and outgoing influences are
explicit in field theories.
\par
{}}. \ It is also apparent that, as assumed in Sect. 4, the natural
frequency $\omega _{o}$, which becomes $\gamma \omega _{o}$ for the moving
particle, is not the measure of some merely internal property of the
particle. \ This is sufficiently evidenced by the de Broglie wave, which is
itself of frequency $\gamma \omega _{o}$, and reveals its presence
externally in diffraction, interference and quantization. \ Conservation
then demands that energies passing to and from the particle be commensurate,
and thus the standing wave (\ref{model}).

ESR assumes that all influences in Nature propagate at velocity $c$.\ \ If
some particle or force were shown to respond at a velocity differing from $c$%
, this would constitute as much an exception to ESR as it would to LSR%
\footnote{%
We do not ignore the nonlocality implied by the formalism of SQM or the
apparent confirmation of that nonlocality in dynamic Bell's experiments,
notably those of Aspect et al \cite{aspect} and Weihs et al \cite{weihs}. \
We are content to rely (here at least) on the "peaceful coexistence" between
SR and SQM declared by Shimony \cite{shimony} \ from the circumstance that
the nonlocality claimed does not permit superluminal signalling.}. \ Massive
particles do not move at that velocity, but it is implicit in ESR that the
influences by which these particles interact do develop between and through
the particles at velocity $c$. \ There is an analogy here with refraction. \
From interference between the incident (free space) wave, and reradiation
from moments induced by that wave, the transmitted wave acquires a phase
velocity that may be greater or smaller than $c$. \ Nonetheless, any change
in the wave develops through the medium at the velocity $c$.

Let us suppose that some particle, let us say a meson mediating the strong
force, or perhaps we should say, a string linking quarks, were to respond at
velocity $C$ (differing from $c$) to some change in the relationship between
one nucleon and another. \ The composition of incoming and outgoing
influences could then be expressed in the form of Eqn. (\ref{twave1}), that
is as,%
\begin{equation*}
\Psi \left( \mathbf{r},t\right) =[e^{i\left( \omega _{1}t-\mathbf{\kappa }%
_{1}\cdot \mathbf{r}\right) }-e^{i\left( \omega _{2}t+\mathbf{\kappa }%
_{2}\cdot \mathbf{r}\right) }]/2,
\end{equation*}%
(or if not in that form then in some superposition of waves of that form),
where contrary to relations (\ref{wk}) above,

\begin{equation*}
\frac{\omega _{1}}{\kappa _{1}}=\frac{\omega _{2}}{\kappa _{2}}=C\neq c.
\end{equation*}%
From the discussion of the LT in Sect. 4, it is thus apparent that on a
change of inertial frame, this supposed meson or string would suffer
contraction, dilation and a changed simultaneity to a degree differing from
that experienced by other matter, so that contrary to ESR (and as will be
discussed in Sect. 8), the laws of physics would not then be the same in all
inertial frames.

Once it is accepted that in its rest frame a particle comprises, or is at
the focus of, influences oscillating at the natural frequency $\omega _{0}$
and moving inwardly and outwardly at velocity $c$, the exact nature of those
influences becomes of no consequence to the demonstration in Sect. 4. \
These influences could be waves or fields, real, virtual or probabilistic,
continuous or particulate. \ They might be concentrated upon some additional
central body, but if having spatial extension, the material of that body
would be of such a nature that influences propagating through the material
did so at velocity $c$. \ Whatever the case, the moving "wave" will display
the contraction and dilation described by the LT, and be subject to a
modulation consistent with the predicted loss of simultaneity.

The plausibility of this model is evidenced by the physically consistent
provenance it provides for the de Broglie wave. \ As to the infinity at the
core of the model, it must be supposed that any viable wave form would
exhibit some asymmetry avoiding such a singularity.

\section{A preferred frame?}

We have supposed, with Lorentz, a preferred frame of reference - the frame
in which the speed of light is not only observed to be, but in fact is, the
same in all directions. \ If, as seems the case, this frame is undetectable,
it might be argued from considerations of economy that it should play no
part in the description of SR. \ Einstein stated:

\begin{quotation}
The introduction of a luminiferous aether will prove to be superfluous
inasmuch as the view here to be developed will not require an "absolutely
stationary space" provided with special properties \cite{einstein1}.
\end{quotation}

Yet space does have at least one "special property" essential to ESR and
assumed implicitly by Einstein. \ Unlike projectiles and massive particles,
one photon never overtakes another. \ Photons pass through space at a common
rate of progress, and their passage is thus similar in this respect to the
propagation of a wave through an elastic medium, where the velocity of the
disturbance is determined only by the nature of the wave and the properties
of the medium. \ 

Einstein suggested that the photon must be an "autonomous entity" requiring
no supporting medium (see Kostro \cite{kostro}, pp. 37 and 94), but it might
be asked why such an unconstrained object should confine itself to a
particular rate of progress. \ Moreover, photons not only propagate as
waves, they have the oscillatory characteristics of waves. \ So too does
matter. \ In the absence of some more plausible explanation, there would
seem ample grounds for suspecting the light-carrying medium supposed by
Lorentz.

Following the completion of GR, Einstein relaxed his stance against the
aether, but not to the extent of admitting that such a medium could
constitute a preferred frame (see generally Kostro \cite{kostro}). \ Yet the
argument for a preferred frame seems even stronger than that for an aether.
\ It may be possible to imagine some reason other than a light-carrying
medium for the photon's fixed rate of progress, perhaps the involvement in
some way of the surrounding Universe. \ But it is the fixed rate of
progress, not how it is enforced, that implies the existence of a frame of
reference.

The argument becomes particularly insistent once the LT is explained from
the wave characteristics of matter. \ There are practical and conceptual
difficulties in presenting a wave based explanation of the LT without
supposing such a frame. \ Consider, for instance, the frequencies $\omega
_{1}$ and $\omega _{2}$ of Eqn. (\ref{wk}). \ No matter how asymmetrical the
Universe might be, it must be possible at least in principle to attain a
velocity in any direction we choose for which one of these frequencies, ($%
\omega _{1}$ or $\omega _{2}$) is the greater, and in the opposite direction
for which the other is the greater, so that between these velocities, there
must exist an inertial frame in which $\omega _{1}=$ $\omega _{2}=\omega
_{0} $. \ There must therefore exist, at least locally, a preferred frame.

These arguments appear as applicable in GR as in SR. \ However, in GR, the
distribution of matter and radiation now hints at the location of such a
frame. \ It would be the frame of the galaxies and of\ the cosmic microwave
background - the frame from which spacetime itself is spatially isotropic -
as is nearly so from the frame of our own local system. \ 

Nor is the covariance of GR (general covariance) as convincingly agnostic of
the existence and location of this frame as the LT of the special theory. \
It was objected by Kretschmann in the first days of GR that by the
sufficient exercise of mathematical ingenuity any equation of physics might
be expressed in covariant form \cite{kretschmann}. \ Kretschmann thus argued
that neither a relativity principle nor the primacy of Einstein's theory
could be established by covariance per se\footnote{%
Kretschmann's objection has been much discussed, see for instance Norton 
\cite{norton}, Dieks \cite{dieks}, and Giovanelli \cite{giovanelli}.}. \
Einstein conceded the point, whilst arguing for the superiority of his
equation on the basis of simplicity \cite{einstein2}. \ And he could also
have cited of course the compelling\ utility and elegance of a manifest
covariance. \ 

Yet any inference from covariance that there is no preferred frame is
weakened in GR by the existence of a gauge freedom allowing a choice of
coordinates adapted to the symmetries of the problem, and thus in effect, to
the curvature of the metric. \ The very notion of curvature implies a centre
of curvature, and it becomes possible to think of the curvature tensor of
the Universe at large as suggesting, mysterious though the concept may be, a
preferred basis for the consideration of the curvature of the Universe
itself.

Further arguments for a preferred frame have emerged that could not have
been guessed at by Lorentz or Poincar\'{e}. \ It has been argued that such a
frame is implied by the assumption, now orthodox, that the vacuum (the
modern aether) is the seat of the zero point energies of the modes of the
fields. \ This has suggested in turn the possibility of frame-dependent
effects in\ the extreme and perhaps pathological situation at the event
horizon of a black hole (see Winterberg \cite{winterberg}). \ There is also
the hypothesized (and, it has been argued, detectable) Unruh-Davies effect,
according to which an observer, who is accelerating with respect to the
vacuum, experiences a thermal bath and consequent temperature rise from
radiative effects (see, for example, Crispino et al \cite{crispino}).

However, it is not essential to the case for Lorentzian relativity presented
here that this\ preferred frame have any particular location or that it be
capable of detection. \ Indeed, there is reason to believe, as the next
section may show, that this frame lies beyond all possibility of detection.
\ But once the LT is explained, not from a change in the properties of
space, but from changes in the wave-constituted material occupying that
space, there must be in consequence one particular frame in which the
underlying waves develop at the same speed in every direction. \ 

If undetectable, this frame can play no essential role in the \textit{%
mathematical} description of SR or GR. \ But with due respect to the friar
of Occam, the dismissal of this frame as "superfluous" has had the effect of
discouraging, if not entirely suppressing, lines of enquiry that imply its
existence.

\section{Lorentz invariance}

While there would seem little doubt that the laws of physics are the same
for all observers, one might ask why this should be so. \ In the context of
ESR this satisfying situation would seem fortuitous. \ However, a
consideration of the wave nature of matter may suggest at least a partial
explanation for this invariance.

Notice firstly the unusual effects that would result if the LT were not
exactly as it is, for instance if the Lorentz factor $\gamma $ was not $%
(1-v^{2}/c^{2})^{-1/2}$ but had some other value $\gamma ^{a}$ ($a$ being a
real number other than unity). \ Consistently with relations (\ref{wk}), a
wave travelling at any desired velocity $v$ could still be constructed, but
as in the case of $a=0$, which is the Galilean transformation, the preferred
frame would then be detectable. \ In the Galilean case, longitudinal
contraction is replaced by transverse contraction that increases with
acceleration away from the preferred frame. \ For other values of $a$,
contraction or expansion (both longitudinal and transverse), time dilation
and simultaneity would be observed to differ according to whether velocity
were increased or decreased with respect to the preferred frame. \ The laws
of physics would not then be invariant, and the Universe would be rather
more curious and less elegant than it is. \ But that alone does not explain
why it is not so.\ 

Consider now, however, the stability of matter. \ The elementary particles
exist in a limited variety of precisely defined characteristic forms. \ Why
that should be so is not apparent, but what is apparent is that they are\
constrained by those forms and by their wave-like nature to combine in a
limited number of ways, as is well illustrated by the Bohr conditions (\ref%
{bohr}). \ Even in matter that we would think of as stationary, underlying
microprocesses are occurring at relativistic speeds, as illustrated well
enough again by the Bohr orbits. \ For these processes to remain undisturbed
by a change of inertial frame, dynamic relationships between particles must
be preserved, including for instance relative velocities, accelerations,
masses and polarizations. \ In other words, the laws of physics must be the
same for all inertial observers, and this is so only at $a=1$.

Considered in this way, Lorentz invariance is not the fortuitous cause, but
rather the inevitable effect, of the manner in which the constituent
elements of matter must persist and combine. \ It is not then the metric
that determines how matter transforms, but the stability of matter that
determines the LT and the observed (Minkowski) metric. \ Of any pre-existing
or primordial metric little can be said, except that the simplest possible
metric, the Euclidean, was that assumed by Lorentz.

\section{Summary}

The effects described by the LT can be explained in their entirety by
changes occurring in matter as it changes inertial frame. \ This is not to
suggest that the LT does not \textit{describe} a transformation of space and
time. \ But what the LT describes are changes in what is observed, and in
the Lorentzian approach offered here, what is observed by an accelerated
observer is essentially an illusion induced by changes in that observer. \ 

This view relies crucially on the conclusion reached in Sect. 1 that the LT
does not involve actual physical change in the properties of space. \ But
once that conclusion is reached, it becomes apparent that there is something
elusive in Einstein's theory, and that it is the Lorentzian approach that
better explains the origin of the contraction, dilation and loss of
simultaneity described by the LT.

Once the LT is explained from the wave characteristic of matter a good deal
else becomes apparent. \ The de Broglie wave is seen to be a modulation
rather than an independent wave, thus explaining the superluminality of this
wave, its significance in the Schr\"{o}dinger equation, and its roles in
determining the optical properties of matter and the dephasing that
underlies the relativity of simultaneity.

Einstein's bold assertion that the laws of physics must be the same for all
observers revealed the elegance of SR and something indeed of the elegance
of the universe itself. \ It is suggested nonetheless that it is a
Lorentzian approach that will provide the deeper understanding of the
physical meaning of SR.

\medskip \bigskip

\textbf{Acknowledgements}

The author thanks Will Nelson and Larry Hoffman for encouragement and
helpful suggestions, and the anonymous reviewers for drawing attention to
relevant and interesting issues arising in the context of general relativity.

\bigskip \smallskip

*email:\ danjune@bigpond.net.au

\end{document}